\documentclass{PoS}
\usepackage{amsmath}

\title{Searches for DM signals from the Galactic Centre region with H.E.S.S.}

\ShortTitle{Searches for DM signals from the Galactic Centre region with H.E.S.S.}

\author{\speaker{Daniil NEKRASSOV}\\%
        Max-Planck Instute for Nuclear Physics, P.O. Box 103980, D 69029 Heidelberg, Germany\\             E-mail: \email{daniil.nekrassov@mpi-hd.mpg.de}}

\author{for the H.E.S.S. Collaboration\\%
        http://www.mpi-hd.mpg.de/hfm/HESS/}

\abstract{The H.E.S.S. experiment, an array of four Imaging Atmospheric Cherenkov Telescopes, observes the Galactic Centre (GC) region in the very high-energy (VHE, E > 100 GeV) domain since 2003. The GC is believed to be the region with the highest Dark Matter (DM) density in our Galaxy, thus making it one of the primary targets for VHE gamma-ray observations, looking for signals from annihilation or decay of DM particles. The interpretation of the collected H.E.S.S. data with regard to possible DM signals is however complicated by the presence of several astrophysical sources of VHE gamma-ray radiation in this region. The current status of the search for DM signals from the GC region is reviewed.}

\FullConference{Identification of Dark Matter 2010\\
                 July 26 - 30 2010\\
                 University of Montpellier 2, Montpellier, France}

\begin{document}

\section{Introduction}

H.E.S.S. (High Energy Stereoscopic System) is an array of four Imaging Atmospheric Cherenkov Telescopes (IACTs), located in the Khomas Highlands of Namibia at an altitude of 1800 m above sea level. Each telescope comprises 382 round mirrors, mounted in a Davies-Cotton design \cite{DaviesCotton:57} with a total mirror area of $\sim 107$ m and a focal length of $15.28$ m \cite{HESS:03}. The mirrors are collecting Cherenkov light, which is emitted by charged constituents of a particle cascade, initiated by the interaction of primary $\gamma$-rays in the Earth's upper atmosphere. The light is subsequently reflected onto a camera, that is sensitive and fast enough to detect such weak Cherenkov flashes of 10 ns duration. The cameras consist of 960 highly sensitive photo-multiplier tubes (PMTs), which provide the needed sensitivity and time resolution. In order to reduce light losses, Winston cones are installed in front of each PMT, focusing the incident light into the active volume of PMTs. The angular size of each PMT is $0.16^{\circ}$ \cite{CAMERA:04}, thus the total field of view of a H.E.S.S. camera accounts to $5^{\circ}$, adapted for observations of extended $\gamma$-ray sources. Thanks to the stereoscopic technique allowing for an accurate reconstruction of the direction and energy of the primary gamma-rays as well as for an efficient rejection of the background induced by cosmic-ray interactions, the energy threshold is about 100 GeV at zenith and the angular resolution is around $0.1^{\circ}$ per $\gamma$-ray. The point source sensitivity is $2 \times 10^{-13} \ \mbox{cm}^{-2} \ \mbox{s}^{-1}$ above 1 TeV for a $5 \sigma$ detection in 25 hours \cite{Crab:06}.\\

The H.E.S.S. experiment participates in the indirect search for DM, which most probably consists of a new type of particles. Such particles are predicted, e.g., by currently existing theories beyond the Standard Model (SM) of particle physics (see, e.g., \cite{Bertone:05} and references therein), like the Supersymmetry (SUSY). In SUSY, the corresponding WIMP (Weakly Interacting Massive Particle) candidate for DM is the lightest neutral particle, most commonly the neutralino $\chi_{0}$, with the mass $\mbox{m}_{\chi}$ between $\sim 100$ GeV and a few TeV. The self-annihilation (or decay in models, where the R-parity is not conserved \cite{Bertone:05}) of such WIMPs, mediated by the weak interaction, would create SM particles in the final state, also including $\gamma$-rays, produced mostly in hadronization processes. Since the spectrum of such $\gamma$-rays would extend up to the WIMP mass, these should be in principle detectable in the VHE domain. For the case of self-annihilations, the expected $\gamma$-ray flux from a certain sky region can be calculated using the following relation:

\begin{equation}
	\frac{d\Phi(\Delta \Omega, E_{\gamma})}{dE_{\gamma}} = \frac{1}{4\pi} \frac{\langle \sigma v \rangle}{\mbox{m}_{\chi}^{2}} \frac{dN}{dE_{\gamma}} \times \bar J \Delta \Omega, \ \mbox{where} \ \bar J = \frac{1}{\Delta \Omega} \int_{\Delta \Omega} \int_{\mbox{line-of-sight}} \rho^2(r) ds.
\label{eq:DM_Flux}
\end{equation}
 
 Here, $\langle\sigma v\rangle$ is the velocity-weighted annihilation cross-section of the corresponding DM candidate and $\frac{dN}{dE_{\gamma}}$ is the spectrum of $\gamma$-rays that can be expected from such annihilations. Both terms entirely depend on the underlying particle physics model. $\bar J$ is the averaged line-of-sight integral of the squared DM density $\rho(r)$ and thus represents the astrophysical input in this relation. The latter term depends on the selected observation targets, of which one of the most promising is the Galactic Centre (GC) region, since to date, it is the closest confirmed region of high DM density (see \cite{Springel:08}, \cite{Diemand:08} for results from recent N-body simulations). Deep VHE $\gamma$-ray observations were performed by H.E.S.S. in this region and various studies of these data in the DM context are described in the following.

\section{H.E.S.S. results on the GC region}

The GC region has been observed by the H.E.S.S. experiment since its operation start in 2003. Together with other IACT experiments, H.E.S.S. has discovered a bright VHE $\gamma$-ray point source HESS J1745-290 at the position of the supermassive black hole (SMBH) Sgr A* in 2004 \cite{HESSGC:04}. Using 50 hours of observations, H.E.S.S. provided the most precise measurement of the energy spectrum of this source, which could be described as a power-law with an index $\Gamma = 2.25 \pm 0.04 (\mbox{stat}) \pm 0.1(\mbox{sys})$. As the origin of the observed emission, the SMBH Sgr A* itself, a nearby supernova remnant Sgr A East and later, the newly discovered pulsar wind nebula (PWN) G359.95-0.04 \cite{PWNGC:07}, were proposed, which could all account for the observed characteristics of the source. On the other hand, DM interactions as the main contribution of this emission were ruled out \cite{HESSGCDM:06}, since the observed spectrum does not follow common parametrizations of $\gamma$-ray spectra, created in DM annihilations, and also extends well beyond a few TeV, thus leading to unnaturally high WIMP masses, if the observed emission would be explained in a DM context (see also \cite{Moulin:07}). \\

Thanks to the continuous observations of the GC region by H.E.S.S. in the following years, the morphological and spectral properties of HESS J1745-290 could be studied in more details. While the SNR Sgr A East can now be excluded as the source counterpart by an improved position measurement, both Sgr A* and the PWN G359.95-0.04 are still in agreement with the position of the source \cite{GCPP:09}. Both candidates can also still reproduce the spectrum, which is now determined to be a power-law with an index $\Gamma = 2.10 \pm 0.04 (\mbox{stat}) \pm 0.1(\mbox{sys})$, having an exponential cut-off of $\mbox{E}_{\mbox{cut}} = 14.7 \pm 3.4$ TeV, making a DM explanation even more unlikely \cite{GCSpectrum:09}. Thus only upper limits on the velocity-weighted annihilation cross-section $\langle\sigma v\rangle$ can be calculated. Such limits exclude SUSY models, comprising a realistic DM candidate, only if a very cusped DM density profile (i.e. for very large $\bar J$), e.g. the Moore profile, is assumed \cite{Ripken:09}. For commonly used DM density profile parametrisations for the Milky Way like Navarro-Frenk-White (NFW) or the Einasto profile, the results are a few orders of magnitude above Moore limits \cite{Ripken:09}. The exact shape of the DM profile within the central few pc of our Galaxy remains speculative, since current N-body simulations cannot reach the needed spatial resolution\footnote{Spatial resolution of current N-body simulations is above 100 pc, see, e.g., \cite{Springel:08}.}, thus upper limits based on observations of the very centre of the GC region are subject to large systematic uncertainties. \\

Apart from the central point source, H.E.S.S. has also discovered an extended region of $\gamma$-ray emission in the GC region \cite{HESSDE:06}. This so-called diffuse emission spatially correlates with giant molecular clouds in this region, thus the interpretation of this emission is that it is due to a local diffusing cosmic ray population, which interact with the surrounding molecular material. A DM interpretation for this emission was not considered, however this measurement, together with H.E.S.S.` results on HESS J1745-290, was used to obtain restrictions of the DM parameter space in a multimessenger context, which are described in the following.

\section{GC observations in a multimessenger context}

One of the advantages of studying DM interactions in the GC region is, due to its proximity, the possibility to relate $\gamma$-ray observations, and observations of electromagnetic radiation in general, with measurements of charged messengers, like $e^{\pm}$ and $p/\bar p$. There exist well studied predictions for the flux of these particles on Earth, which are based on current knowledge about particle acceleration and subsequent propagation in space (see, e.g., \cite{GalProp:09}). Measured deviations from these predictions, like the PAMELA positron anomaly \cite{PAMELA:08}, can thus be interesting in a DM context. Since also the production of charged particles in DM interactions mostly takes place in the GC region\footnote{For completeness it is noted that charged particles from DM interactions could also be produced in nearby DM clumps. In this case, the connection between charged messengers and observations of the GC region is not valid.}, DM models trying to account for the observed anomalies can be tested with observations of the GC throughout the entire electromagnetic spectrum. This has been done, e.g., with regard to the measurements of the $e^{\pm}$ spectra done by FERMI (see \cite{FERMI:10} for latest results) and PAMELA \cite{PAMELA:08}. The regions of the parameter space favoured by DM models accounting for these measurements were tested using radio and $\gamma$-ray observations of the GC region done by H.E.S.S., where both the spectrum of the central point source and the spectrum of the diffuse emission were used as upper limits for a possible DM flux \cite{Meade:09}. Even with such conservative assumptions, the VHE data helped to exclude the regions of the DM parameter space under study for the NFW and Einasto profiles. It is noteworthy, that limits obtained using the $\gamma$-ray flux from the diffuse emission were compatible with those using the GC point source data (see also \cite{Crocker:10} for a similar analysis). Thus $\gamma$-ray observations of the area outside the inner few pc is also capable of providing sensitive upper limits.

\section{Current activities}

\begin{figure}
	\includegraphics[width=0.7\textwidth]{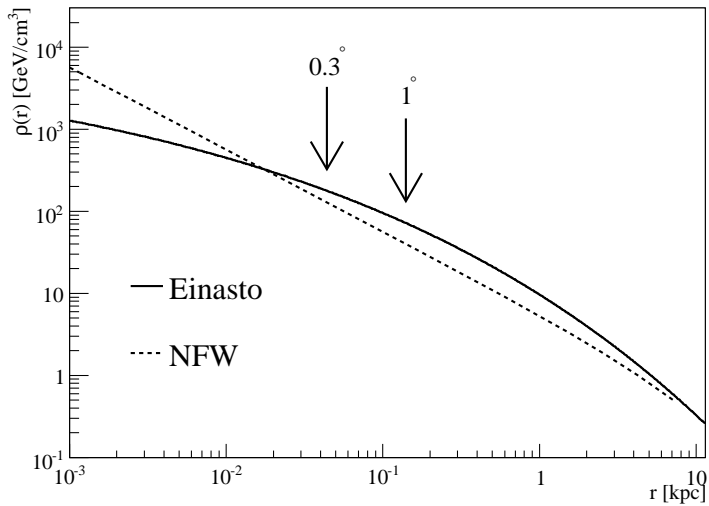}
	\caption{Comparison of the NFW and Einasto DM density profiles. The parameters for the NFW and Einasto profiles are taken from \cite{Pieri:09}. Both profiles are normalized to the local DM density ($\rho_0 = 0.39$~GeV/cm$^{3}$ at a distance of 8.5~kpc from the GC). The region used to extract upper limits by a recent analysis approach, studying the GC DM halo, is indicated by arrows, which are located $0.3^{\circ}$ and $1^{\circ}$ away from the GC. See text for further details.}
	\label{Fig:Profiles}
\end{figure}

Currently more than 100 hours of observation time were spent on the GC region by H.E.S.S.. As far as the central point source results are concerned, no significant improvement of the existing results can be expected, since this would require a notable, and currently unrealistic increase of observation time. However, as was shown above, interesting results can be obtained when studying the GC region outside the point source. Until now, only the band of VHE diffuse emission along the Galactic Plane was used for DM studies. An improvement upon this approach should be achieved by a dedicated analysis of the total GC region of up to several degrees in extension, called GC halo \cite{Schwanke:09}, while excluding astrophysical VHE $\gamma$-ray sources at the same time. This is reasonable, since even when avoiding the very centre, the DM density in the remaining part of the GC halo should still be quite high, see Fig. \ref{Fig:Profiles}. Due to the size of the GC halo, subtraction of the numerous hadronic background inside such an area is a critical issue. 

Currently there exist two procedures to perform such a study: Either new observations of the GC region have to be conducted, together with observations of empty regions of the sky, that could then be used for background subtraction \cite{Schwanke:09}. This method can make use of the total available field of view of H.E.S.S. and cover the region around the GC up to a few degrees. For another approach, a dedicated background subtraction, using control regions from the same field of view, is needed to make use of the existing dataset (see also \cite{Crab:06}). An analysis based on the latter approach has been recently performed, extracting limits on $\langle\sigma v\rangle$ from a circular area around the GC of $1^{\circ}$ extension, while using the remaing field of view for background estimation \cite{GCHalo:11}. As a result, values of $\langle\sigma v\rangle$ above $3 \times 10^{-25} \text{ cm}^{3}\text{ s}^{-1}$ are excluded for the Einasto density profile. It should be emphasized that the systematic uncertainty due to the shape of the DM profile is significantly reduced, since the very centre of the Galaxy is avoided.

\section{Conclusions}

The GC region remains one of the most promising targets for the indirect search for DM. However, in the VHE $\gamma$-ray domain, astrophysical $\gamma$-ray radiation hampers the search for DM signals. Still, interesting constraints on the DM parameter space can be obtained if using the VHE data in a multimessenger context. Currently, analyses of the parts of the GC region without contamination by astrophysical sources are ongoing, promising sensitive contraints on $\langle\sigma v\rangle$ while reducing at the same time the uncertainties due to the shape of the DM density profile.


\begin{thebibliography}{99}
  \bibitem{DaviesCotton:57}
     Davies,~J. and Cotton,~E., \\
     \emph{Journal of Solar Energy Science and Engineering} {\bf 1} (1957) 16

 \bibitem{HESS:03}
	Bernl\"ohr,~K et al., \emph{The optical system of the H.E.S.S. imaging atmospheric Cherenkov telescopes}, \\
	\emph{Astropart. Phys.} {\bf 20} (2003) 111.


	\bibitem{CAMERA:04}
	Aharonian, F et al., \emph{Calibration of cameras of the H.E.S.S. detector},\\
	\emph{Astroparticle Physics}, {\bf 22} (2004) 109 [{\tt astro-ph/0408145}]
	\bibitem{Crab:06}
	Aharonian, F et al., \emph{Observations of the Crab nebula with HESS},\\
	\emph{A\&A}, {\bf 457} (2006) 899 [{\tt astro-ph/0607333}]

	\bibitem{Bertone:05}
	 G.~Bertone et al., \emph{Particle dark matter: evidence, candidates and constraints}, \\
	 \emph{Phys. Rep.}, {\bf 405} (2005) 279 [{\tt hep-ph/0404175}]

	\bibitem{Bergstroem:09}
   L. Bergstr\"om, \emph{Dark matter candidates}, \\
  \emph{New Journal of Physics} {\bf 11} (2009) 105006 [{\tt hep-ph/0903.4849}]

	\bibitem{Springel:08}
   V.~Springel et al., \emph{Prospects for detecting supersymmetric dark matter in the Galactic halo},\\
	\emph{Nature}, {\bf 456} (2008) 73

	\bibitem{Diemand:08}
	J. Diemand et al., \emph{Clumps and streams in the local dark matter distribution},\\
	\emph{Nature}, {\bf 454} (2008) 735
 
	\bibitem{HESSGC:04}
	F. Aharonian et al., \emph{Very high energy gamma rays from the direction of Sagittarius A$^{*}$},\\
	\emph{A\&A}, {\bf 425} (2004) L13 [{\tt astro-ph/0406658}]  
	
	\bibitem{PWNGC:07}
	Q.~D. Wang et al., \emph{G359.95-0.04: an energetic pulsar candidate near Sgr A*},\\
	\emph{MNRAS}, {\bf 367} (2006) 937 [{\tt astro-ph/0512643}]  

	\bibitem{HESSGCDM:06}
	F. Aharonian et al., \emph{HESS Observations of the Galactic centre Region and Their Possible Dark Matter Interpretation}, \emph{PRL}, {\bf 97} (2006) 221102 [{\tt astro-ph/0610509}]  

	\bibitem{Moulin:07}
	E. Moulin, \emph{Indirect dark matter searches with H.E.S.S},\\
	in proceedings of \emph{SUSY 07}, [{\tt astro-ph/0710.2493}]

	\bibitem{GCPP:09}   
	F. Acero et al., \emph{Localising the VHE gamma-ray source at the Galactic Centre},\\
	\emph{MNRAS}, {\bf 402} (2010) 1877 [{\tt astro-ph/0911.1912}]  
	
	\bibitem{GCSpectrum:09}
	F. Aharonian et al., \emph{Spectrum and variability of the Galactic centre VHE {$\gamma$}-ray source HESS J1745-290}, \emph{A\&A}, {\bf 503} (2009) 817 [{\tt astro-ph/0906.1247}]  
	
	\bibitem{Ripken:09}
	J. Ripken et al., \emph{Update on Dark Matter in the Galactic Centre with H.E.S.S.},\\	
	in proceedings of \emph{ICRC 09}
	
	\bibitem{HESSDE:06}
	F. Aharonian et al., \emph{Discovery of very-high-energy {$\gamma$}-rays from the Galactic Centre ridge},\\
	\emph{Nature}, {\bf 439} (2006) 695 [{\tt astro-ph/0603021}]  

	\bibitem{GalProp:09}
	A.~W. Strong et al., \emph{The GALPROP Cosmic-Ray Propagation Code},\\ 
	{\tt astro-ph/0907.0559}

	\bibitem{PAMELA:08}
	Adriani, O. et al., \emph{An anomalous positron abundance in cosmic rays with energies 1.5-100GeV},\\
	\emph{Nature}, {\bf 458} (2008) 607 [{\tt astro-ph/0810.4995}]		
	
	\bibitem{FERMI:10}
	 Fermi LAT collaboration, \emph{Fermi LAT observations of cosmic-ray electrons from 7 GeV to 1 TeV},\\
	 {\tt astro-ph/1008.3999}
   
   \bibitem{Meade:09} 
	P. Meade et al., \emph{Dark Matter interpretations of the $e^{±}$ excesses after FERMI},\\
	\emph{Nuclear Physics B}, {\bf 831} (2010) 178 [{\tt hep-ph/0905.0480}]		

	\bibitem{Crocker:10}
	 R.~M. Crocker et al., \emph{Radio and gamma-ray constraints on dark matter annihilation in the Galactic centre}, \emph{Physical Review D}, {\bf 81} (2008) 63516 [{\tt hep-ph/1002.0229}]		

	\bibitem{Schwanke:09}
	U. Schwanke et al., \emph{Dark Matter Searches with Imaging Atmospheric Cherenkov Telescopes and Prospects for Detection of the MilkyWay Halo}, in proceedings of \emph{PATRAS 09}		
		
	\bibitem{Pieri:09}		
	 L. Pieri et al, \emph{Implications of High-Resolution Simulations on Indirect Dark Matter Searches},\\
	{\tt astro-ph/0908.0195}
	
	\bibitem{GCHalo:11}
	A. Abramowski et al, \emph{Search for a Dark Matter Annihilation Signal from the Galactic Center Halo with H.E.S.S.},\\
	\emph{PRL}, {\bf 106} (2011) 161301 [{\tt astro-ph/1103.3266}]

		
\end{thebibliography}
\end{document}